\documentclass[aps,prb,twocolumn,groupedaddress]{revtex4}  
\usepackage{graphicx}

\begin{document}   
\newcommand{\be}{\begin{equation}}  
\newcommand{\ee}{\end{equation}}  
\newcommand{\bea}{\begin{eqnarray}}  
\newcommand{\eea}{\end{eqnarray}}  
\newcommand{\nt}{\narrowtext}  
\newcommand{\wt}{\widetext}    

\title{Coulomb interacting Dirac fermions in disordered graphene}   

\author{D. V. Khveshchenko}   

\affiliation{Department of Physics and Astronomy, University of North Carolina, Chapel Hill, NC 27599}    

\begin{abstract}  
We study such experimentally relevant characteristics 
of the Coulomb interacting Dirac quasiparticles in disordered graphene
as the quasiparticle width and density of states that can be probed by
photoemission, magnetization and tunneling measurements. 
We find that an interplay between the unscreened
Coulomb interactions and pseudo-relativistic quasiparticle kinematics
can be best revealed in the ballistic regime, whereas in the diffusive limit
the behavior is qualitatively similar
to that of the ordinary 2DEG with parabolic dispersion.
\end{abstract}    
\maketitle 
 
The recent advances in microfabrication of graphitic monolayers \cite{geim} 
have made it possible to test and confirm the earlier theoretical predictions of anomalous, 
relativistic-like, kinematic properties of the electronic states in graphene \cite{semenoff}.
Thus far, both experimental and theoretical studies have 
been primarily focusing on the unusual (magneto)transport phenomena which, 
for their most part, can be described in terms of non-interacting Dirac 
quasiparticles propagating ballistically in the bulk or in finite geometries.

However, it has long been recognized that a (nearly) degenerate 
two-dimensional semimetal such as graphene might provide a unique 
playground for studying the effects of the Coulomb interactions, 
because the latter are expected to remain essentially unscreened 
\cite{guinea1}. 

Among the previously discussed manifestations of the Coulomb correlations is a possible
opening of the interaction-induced excitonic gap at sufficiently strong Coulomb couplings \cite{dvk1}.
Alternatively, such a gap can 
also be generated by a magnetic field, which phenomenon represents the Dirac counterpart 
of FQHE in the conventional 2DEG
with parabolic electron dispersion, as was pointed out in Ref.\cite{dvk2} (see also Ref.\cite{gorbar} for the
references to the earlier studies of a related phenomenon of "magnetic catalysis" in abstract field-theoretical setting).

Should a spectral gap develop, it would be exhibited by the conductivity and other transport characteristics.
However, apart from the recent observation \cite{kim} of a complete lifting of the 
four-fold degeneracy of the $n=0$ Landau level (which phenomenon
would indeed be consistent with the scenario of a field-induced gap opening, resulting 
in a spontaneous breakdown of the sublattice symmetry \cite{dvk2,gorbar}), no conclusive  
evidence of such a behavior has yet been found. 

Also, if proven to be of a genuine bulk nature (as opposed to being due to magnetic 
impurities, edges and/or structural defects), the previously reported weak, albeit robust, 
ferromagnetism in pyrolytic graphite \cite{kopelevich1}
could be indicative of a possible instibility towards a (weakly) ferromagnetic excitonic state 
with unequal gaps for the spin-up and spin-down electrons \cite{dvk1}. 

Obviously, a further experimental work is needed in order to 
accertain a real status of the scenario of a latent excitonic insulator
proposed in Ref.\cite{dvk1}, as well as contrasting it with such
alternative predictions as that of the Stoner instability 
resulting in a fully polarized ferromagnetic state \cite{guinea2}.

In light of this uncertainty, in the present Communication we focus on the effects of 
the moderately strong Coulomb correlations which might not 
be powerful enough to generate a finite gap in the Dirac spectrum. 
As we demonstrate below, even in this case the quasiparticle properties can be affected 
in a number of experimentally relevant ways.
To that end, we study the quasiparticle width and density of states
in both, the ballistic and diffusive regimes, and contrast the results against
those pertaining to the ordinary 2DEG with parabolic dispersion.

An extensive experience gained in the course of the previous studies of the conventional 2DEG 
suggests that such transport characteristics as the longitudinal and Hall DC conductivities 
may not provide the best means of revealing the Coulomb correlations. A general reason is that the two-particle 
response functions probed by transport measurements appear to be only weakly affected by such 
correlations due to a routine cancelation between the 
(potentially, large) fermion self-energy and vertex corrections. In that regard, 
a greater insight into the physics of interacting Dirac 
fermions can be provided by various single-particle probes, including 
photoemission, tunneling, and magnetization measurements. 

The low-energy properties of graphene are governed by the 
electronic states in the vicinity of one of the two inequivalent nodal points ($\alpha=1,2$).
Such states can be described by the Dirac Hamiltonian \cite{semenoff,guinea1}  
\bea 
H=iv_F\sum_{\alpha=1,2}\int_{\bf r}\Psi^\dagger_\alpha[{\hat \sigma}_x\nabla_x+(-1)^\alpha
{\hat \sigma}_y\nabla_y]\Psi_\alpha~~~\\
+{v_F\over 4\pi}\sum_{\alpha,\beta=1,2}\int_{\bf r}\int_{\bf r^\prime}{
\Psi^\dagger}_\alpha({\bf r}^\prime)\Psi_{\alpha}({\bf r}^\prime)
{g\over {|{\bf r}-{\bf r}^\prime|}} {\Psi^\dagger}_\beta({\bf r})\Psi_\beta({\bf r})\nonumber  
\eea 
where $v_F$ is the Fermi velocity, $g_0=2\pi e^2/\epsilon_0v_F\sim 3$ is the bare value of the 
dimensionless Coulomb coupling, and ${\hat \sigma}_i$ is a triplet of the Pauli 
matrices acting in the space of (pseudo)spinors $\Psi_\alpha=(\psi_\alpha(A),\psi_\alpha(B))$ 
composed of the values of the electron wave function on the $A$ and $B$ sublattices
of the bipartite hexagonal lattice of graphene.

The effects of the Coulomb interactions on the fermion propagator and interaction function 
are encoded in the fermion self-energy $\Sigma$ and polarization operator $\Pi$
\bea
{\hat G}^R_\alpha(\omega,{\bf p})^{-1}=(\omega+\mu){\hat 1}-v_F({\hat \sigma}_xp_x+(-1)^\alpha
{\hat \sigma}_yp_y)\nonumber\\
+{\hat \Sigma}^R(\omega,{\bf p}),~~~~ V^R(\omega,{\bf q})=[{q\over g_0}+\Pi^R(\omega,{\bf q})]^{-1}
\eea 
In the one-loop approximation, the former is given by the expression  
\bea  
{\hat \Sigma}^R(\epsilon,{\bf p})=~~~~\\
=\int{d\omega\over (2\pi)}\sum_{\bf q} [ImV^A(\omega,{\bf q}){\hat G}^R(\epsilon+\omega,{\bf p}+{\bf q})
\coth{\omega\over 2T}\nonumber\\ 
-V^A(\omega,{\bf q})Im{\hat G}^R(\epsilon+\omega,{\bf p}+{\bf q})
\tanh{\epsilon+\omega\over 2T}]\nonumber
\eea  
The linear-in-momentum term in $Re{\hat \Sigma}^R(\epsilon,{\bf p})$ gives rise to a renormalization 
of the running Coulomb coupling $g(\omega)$ described by the RG equation derived in Ref.\cite{guinea1} 
\be 
{dg(\omega)\over d\ln(\Omega/\omega)}=-{1\over 8\pi}g^2(\omega)
\ee
The solution $g(\omega)\approx{g_0/[1+(g_0/8\pi)\ln(\Omega/\omega)]}$, where $\Omega$ is 
an upper cutoff of order the electronic 
bandwidth, shows that the effective Coulomb coupling slowly decreases with decreasing energy.
 
At a finite temperature $T$, chemical potential $\mu$ or elastic quasiparticle width $\gamma$ the RG 
flow described by Eq.(4) terminates below the energy scale $\sim max[T, \mu,\gamma]$.    

At $T>0$ and/or in the presence of disorder, a functional form of the fermion 
polarization operator becomes quite prohibitive. However, in the ballistic limit and 
near half-filling ($\mu\approx 0$), it can still be approximated as follows
\bea
\Pi^R(\omega, {\bf q})\approx{1\over 4v_F}
{{\bf q}^2\over {\sqrt {v^2_F{\bf q}^2-(\omega+i0)^2}}},~~~Q_+\gg T\nonumber\\
\approx{2T\ln 2\over \pi v_F}(1-{\omega\over {\sqrt {(\omega+i0)^2-v^2_F{\bf q}^2}}}),
~~~Q_+\ll T~~~
\eea
where $Q^2_+=\omega^2+v^2_F{\bf q}^2$. 

In Eq.(5), the first expression is the temporal component of the Lorentz-invariant free 
fermion polarization bubble
computed at $T=0$, whereas the second one exhibits the (well known in high energy physics) 
phenomenon of "thermal 
Debye screening" and concomitant Landau damping due to thermally excited quasiparticles.
 
A straightforward analysis of Eq.(3) shows that the inelastic quasiparticle 
width defined as $\Gamma(\epsilon,{\bf p})=Im Tr{\hat \Sigma}^R_{in}(\epsilon,{\bf p})$ 
exhibits a strong "light-cone" singularity, akin to that previously 
encountered in the studies of the normal quasiparticles 
in $d$-wave superconductors where the commonly 
quoted $T^3$-behavior of the inverse quasiparticle lifetime represents a rough estimate that 
is only applicable to the thermal quasiparticles with energies and momenta
$\epsilon\sim vp\sim T$ (see Ref.\cite{paaske}). 

When evaluated to the lowest order in the Coulomb coupling, $\Gamma(\epsilon,{\bf p})$ 
appears to be discontinuous 
at $\epsilon=v_Fp$ and singular at $\epsilon, v_Fp\to 0$: 
\bea 
\delta\Gamma(\epsilon,{\bf p})\sim g^2\theta(P^2_-)P_+,~~~ P_+ > T
\nonumber\\
\sim g^2\theta(P^2_-){T^2\over P_+},~~~ P_+ < T
\eea 
where $P^2_{\pm}=\epsilon^2\pm v_F^2{\bf p}^2$, and $\theta(x)$ is the Heaviside step-function.
 
In both regimes, the dominant contribution comes from 
the transferred momenta of order $\sim P_+$ due to the long-ranged 
nature of the Coulomb 
coupling (for a screened interaction, the last line in Eq.(6) would be replaced with
$\sim (T^3/P_+)^{1/2}$).

At ${p}=T=0$ the linear energy dependence of the quasiparticle 
width was predicted in Refs.\cite{guinea1}. It is worth noting, however, that in the case 
of bulk graphite discussed in Ref.\cite{guinea1} the linear dependence would
only hold at the momenta higher than the inverse inter-layer separation $1/d$, while for $q\lesssim 1/d$ the screened interaction potental becomes less singular, $V({\bf q})\approx g_0(d/{q})^{1/2}$. 

By contrast, in the case of graphene Eq.(6) would indeed hold all the way down to the energies
$\sim max[T, \mu,\gamma]$, if it were not for the higher order corrections. 
In order to estimate their effect, we recalculate
Eq.(3) with the RPA-dressed interaction function accounting for the fermion polarization (5), 
which procedure yields the result
\bea 
\Gamma(\epsilon,{\bf p})\sim \theta(P^2_-){{P^2_-}
\over P_+}\ln g,~~~ P_+ \gtrsim gT\nonumber\\
\sim \theta(P^2_-)(g{{P^2_-}T\over P_+})^{1/2}\ln{P_+\over T},~~~ T\lesssim P_+ \lesssim gT
\eea 
while for $P_+\lesssim T$ the logarithmic factor (which would have been absent altogether 
for any interaction less singular than Coulomb) disappears from Eq.(7).
In contrast to Eq.(6), the result (7) remains continuous at the threshold $\epsilon=v_Fp$. 

In the presence of potential disorder, the low-energy quasiparticle width is 
dominated by the elastic part of the self-energy. 
A closed expression for the latter can be readily obtained 
in the case of short-range impurities with concentration 
$n_i$ and scattering amplitude $u$
\be 
{\hat \Sigma}_{el}^R(\epsilon,{\bf 0})={n_iu^2(\epsilon+i\gamma)\ln{(\Omega/\epsilon+i\gamma)}
\over {1-u^2(\epsilon+i\gamma)^2\ln^2{(\Omega/\epsilon+i\gamma)}}}{\hat {\bf 1}} 
\ee 
Eq.(8) allows for a self-consistent calculation of the zero-energy quasiparticle width 
$\gamma=Im Tr{\hat \Sigma_{el}}^R(0,0)$, ranging from the Born ($u\to 0$) to the unitarity ($u\to \infty$) limit.  It is worth mentioning, however,
that Eq.(8) would need to be further modified
in the potentially relevant case of the Coulomb impurities \cite{nomura}.
 
Having obtained $\gamma$, one can compute the non-interacting
DOS at the Fermi energy and the corresponding Drude conductivity 
\bea
\nu_0=-{1\over \pi}Im Tr\sum_{\bf p}{\hat G}^R_0(0,{\bf p})
\approx max[{\gamma\over 2\pi v^2_F}\ln{\Omega\over \gamma},{4\mu\over v^2_F\pi}],
\nonumber\\
\sigma_0\approx{e^2\over h}max[{4\over \pi},{\mu\over \gamma}]~~~
\eea
where ${\hat G}_0^R(\epsilon,{\bf p})$ accounts for the impurity-induced 
broadening, but does not include any inelastic scattering.

In order to study a crossover between the ballistic and diffusive regimes, we use the formula
\be
\Pi^R(\omega,{\bf q})={1\over 4v_F}{{\bf q}^2\over {\sqrt {v^2_F{\bf q}^2-(\omega+i\gamma)^2}}-\gamma}
\ee
interpolating between Eq.(5) for $Q_+\gtrsim \gamma$ and the standard diffusive 
expressions, such as $\Pi^R(\omega,{\bf q})=\sigma_0{\bf q}^2/(D{\bf q}^2-i\omega)$, 
for $Q_+\lesssim  \gamma$
(here $D=\sigma_0/\nu_0$ is the diffusion coefficient).

In the ballistic limit ($\epsilon, T\gg\gamma$) the total self-energy is approximately given by the sum 
of Eqs.(7) and (8), whereas in the opposite, diffusive, regime ($\epsilon, T\lesssim \gamma$) the inelastic  
width can be found from a self-consistent equation
\bea
\Gamma(\epsilon,T)=\int^{\infty}_{-\infty}{d\omega\over 2\pi}
[\tanh({\epsilon+\omega\over 2T})-\coth({\omega\over 2T})]\nonumber\\
\times\sum_{\bf q}Im{V^R(\omega,{\bf q})\over Dq^2-i\omega+\Gamma(\epsilon+\omega,T)}
\eea
whose solution behaves as 
\be 
\Gamma(\epsilon, T)\sim {max[\epsilon,T]\over \sigma_0}\ln{\sigma^2_0g^2_\gamma
\gamma\over max[\epsilon,T]}
\ee
where $g_\gamma=g(\omega\sim\gamma)$.

The total quasiparticle width can be deduced from the ARPES data \cite{arpes}.  
Alternatively, it can be inferred from the dHvA experiments and 
(to the extend that the oscillating part of the resistivity
is indicative of the behavior of the single-particle Green function) the SdH ones.
Namely, fitting the magnetization data to the formula
\be
\Delta M(B)\sim {T\mu^2\over eBv^2_F}\sum_{n=1}^\infty
{n\sin({\pi n\mu^2/eB})\over \sinh(2\pi^2nT\mu/eB)}e^{-2\pi n\mu\Gamma(\mu,T)/eB}
\ee
can provide, apart from such a spectacular hallmark 
of the free Dirac kinematics as the geometric (Berry) phase $\pi$ \cite{kopelevich2}, 
a valuable information 
on the energy/temperature dependence of the quasiparticle width.

In the same spirit, one can estimate the Coulomb-controled phase breaking time 
$\Gamma_\phi(T)\sim (T/\sigma_0)\ln\sigma_0$ 
whose temperature dependence 
(familiar from the theory of the conventional 2DEG) can be manifested by magnetoresistance 
associated with the localization corrections to the Drude conductivity \cite{ando}.

Yet another viable experimental probe is provided by tunneling measurements.
Previous studies have been primarily concerned with the behavior of the electronic DOS
in the vicinity of strong potential impurities \cite{dos1}. 
However, despite offering a greater experimental observability of such prominent 
features as a resonant peak at (or close to, if the particle-hole symmetry is 
broken by subdominant terms
in Eq.(1)) zero energy, the near-impurity DOS appears to be highly non-universal
and, therefore, reveals more information about the impurity potential itself 
than about the Coulomb correlations in the host electronic system.
Notably, a typical plot of the near-impurity DOS \cite{dos1} appears to be very similar 
to that obtained in the case of a $d$-wave superconductor 
(see, e.g., \cite{atkinson} and references therein)
where the Coulomb interactions would be completely screened out by the condensate.
  
In view of the above, in what follows we concentrate on the 
bulk DOS, the first interaction correction to which
is given by the expression 
\bea 
\delta\nu(\epsilon)={1\over \pi}\sum_{\bf p}Im Tr[{\hat G}_0^R(\epsilon,{\bf p})]^2
{\hat \Sigma}^R(\epsilon, {\bf p})=\nonumber\\
\sim -g\epsilon\ln{\Omega\over\epsilon},~~~\epsilon, T\gg \gamma\nonumber\\
\sim -{\nu_0\over \sigma_0}\ln{\gamma\over \epsilon}
\ln{{\tilde \gamma}\over \epsilon},~~~~\epsilon, T \lesssim \gamma~~~~ 
\eea
where ${\tilde \gamma}=\gamma\sigma^4_0g^4_\gamma$.
In the ballistic regime, the correction to the bare (linear) DOS features an additional 
(as compared to the case of the conventional disordered 2DEG \cite{dos2})
logarithmic factor due to the aforementioned 
kinematic "light-cone" singularity, while in the diffusive limit one obtains the same 
diffusion-related (double-$\log$) enhancement, as in the standard case.
 
Associated with the DOS correction (14), there 
are the Altshuler-Aronov-type contributions to such 
observables as specific heat and quasiparticle conductivity which, 
unlike their weak-localization counterparts \cite{ando}, 
can not be readily suppressed by external in-plane magnetic field. 

Beyond the leading approximation, one finds an interference between the Coulomb interactions and disorder, 
which further modifies the behavior of the idealized (clean and non-interacting) Dirac fermion system.  
Given the large bare strength of the Coulomb interaction, the higher order terms might contribute significantly, thus prompting one to employ an adequate non-perturbative technique. 

To that end, we make use of the tunneling action method of Refs.\cite{nazarov}. 
Adapting this approach to the case of 
graphene, we cast the tunneling DOS in the form
\be
\nu(\epsilon)\approx -{1\over \pi}Im Tr\int^\infty_{-\infty}
{\hat G}^R_0({\bf 0},t)e^{-S(t)+i\epsilon t}{dt}
\ee 
The disorder-averaged real-space/time Green function 
${\hat G}_0^R({\bf 0}, t)\propto e^{-\gamma t}/t^2$ is computed
in the absence of the Coulomb interactions, while the 
latter are incorporated through the (imaginary part of) the action 
\bea
S(t)=\int{d\omega\over 4\pi}\coth{\omega\over 2T}\sum_{\bf q}Im V(\omega,{\bf q})\nonumber\\
\times\int^t_0dt_1\int^t_0dt_2e^{-i\omega(t_1-t_2)}<e^{i{\bf q}({\bf r}(t_1)-{\bf r}(t_2))}>
\eea 
which describes the spreading of the excess charge associated with an act of 
tunneling into the graphene sample from, e.g., the STM tip.

In the path-integral language, the averaging in Eq.(16) is carried out
over all the quasiparticle trajectories ${\bf r}(t)$ contributing to the tunneling amplitude
\cite{nazarov}. In the ballistic regime, one obtains $<e^{i{\bf q}
({\bf r}_1-{\bf r}_2)}>\approx 1$, 
whereas in the diffusive limit $<e^{i{\bf q}({\bf r}_1-{\bf r}_2)}>\approx 
e^{-D{\bf q}^2}$. 

To facilitate a direct contact with experiment, we evaluate the tunneling conductance
\bea
G(V,T)\sim {d\over dV}\int_0^\infty[n(V+\epsilon)-n(\epsilon)]\nu_{FL}(V+\epsilon)
\nu(\epsilon)d\epsilon
\nonumber\\
\propto \int d\epsilon{\nu(\epsilon)\over T\cosh^2(V+\epsilon/2T)}~~~
\eea 
where $\nu_{FL}(\epsilon)\approx const$ is the electron DOS of the normal (Fermi-liquid-like) STM tip
biased at a voltage $V$.

In the ballistic regime ($V, T \gg \gamma$) Eq.(16) yields
\bea
S(t)\approx{g_0^2\over (4\pi)^2}\ln(\Omega t), ~~~g_0\ll 1\nonumber\\  
\approx{1\over \pi^2}\ln(\Omega t)\ln[{8\pi\over e}\ln(\Omega t)], ~~~g_0\gg 1
\eea
thereby resulting in the approximate power-law behavior 
\be
G(V,T)\propto max[V, T]^{1+\eta}
\ee
At weak coupling, the "zero-bias anomaly" (19) features 
a purely algebraic behavior with the anomalous exponent $\eta={g^2_0/(4\pi)^2}$.
In contrast, at strong bare coupling the energy dependence of $g(\omega)$ 
gives rise to an approximate power-law decay where
the effective exponent $\eta(V,T)$ deviates slowly (only as $\sim\ln\ln\Omega/max[V,T]$)
from the universal value $\eta=(1/\pi^2)\ln(8\pi e)\approx 0.43$ 
attained at $max[V,T]=\Omega$.

In the diffusive regime ($T, V\lesssim \gamma$), the running coupling $g(\omega)$ levels off at the 
value $g_\gamma$, and the counterpart of Eq.(18) reads 
\bea 
S(t)\approx{1\over (4\pi)^2\sigma_0}\ln(t{\tilde\gamma})[\ln(t\gamma)+O(1)],~~~ 1/\gamma <t<1/T,\nonumber\\
\approx{1\over (4\pi)^2\sigma_0}\ln{{\tilde \gamma}\over T}[\ln{\gamma\over T}+2Tt],~~~t> 1/T~~~~
\eea
As a result, for $\mu\gg\gamma$ (or $\sigma_0\gg 1$) there exists an 
interval ${\sqrt \sigma_0}<\ln(\gamma/max[V,T])<\sigma_0$ where one obtains the dependence 
similar to that of 
the conventional disordered 2DEG \cite{nazarov}
\be
G(V,T)\propto\nu_0\exp(-{1\over 16\pi^2\sigma_0}\ln({\gamma\over 
max[V,T]})\ln({{\tilde \gamma}\over max[V,T]}))
\ee
At still lower biases and/or temperatures 
($max[V,T]<\gamma e^{-4\pi^2\sigma_0}$) the conductance 
resumes a linear dependence
\be
G(V,T)\propto{e^{4\pi^2\sigma_0}\over {\sigma_0}^{1/2}g^2_\gamma}max[V,T]  
\ee
reminiscent of the non-interacting DOS, but with a completely different
prefactor. 
In contrast to the intermediate asymptotic regime (21) that can only occur in the strongly metallic 
case ($\mu\gg\gamma$), 
the linear dependence (22) might be expected to set in at the lowest biases/temperatures for an 
arbitrary electron density. It is worth emphasizing, however, 
that, unlike in the case of the conventional 2DEG, the bare 
DOS of graphene is entirely due to disorder at 
low electron densities ($\mu\lesssim\gamma$).

In summary, we analyzed the effects of moderately strong
Coulomb interactions on the Dirac quasiparticle excitations in graphene.
Taken at their face values, the above results for the quasiparticle width and DOS
suggest that the Dirac physics can be best revealed in the 
ballistic regime (see Eqs.(7) and (19)), while the diffusive dynamics of this system
(see Eqs.(12) and (21,22)) appears to be deceptively similar to that of the conventional 2DEG. These predictions of both, novel and mundane, features 
can be tested in future experiments on photoemission, tunneling, and magnetization measurements.
 
This research was supported by NSF under Grant DMR-0349881.

\end{document}